# Adaptive Queue Prediction Algorithm for an Edge-Centric Cyber–Physical System Platform in a Connected Vehicle Environment

## Paper number: 18-06586

**Mizanur Rahman***
**Ph.D. Candidate**
Glenn Department of Civil Engineering, Clemson University
351 Flour Daniel, Clemson, SC 29634
Tel: (864) 650-2926, Fax: (864) 656-2670
Email: mdr@clemson.edu

**Mashrur Chowdhury, Ph.D., P.E., F. ASCE**
**Eugene Douglas Mays Endowed Professor of Transportation and**
**Professor of Civil Engineering, Professor of Automotive Engineering and**
**Professor of Computer Science**
Glenn Department of Civil Engineering, Clemson University
216 Lowry Hall, Clemson, South Carolina 29634
Tel: (864) 656-3313, Fax: (864) 656-2670
Email: mac@clemson.edu

**Anjan Rayamajhi, Ph.D. Student**
Department of Electrical and Computer Engineering,
314 Flour Daniel, Clemson University, SC 29634
Tel: (864)-633-8659, Fax: (864)-656-0145
Email: arayama@g.clemson.edu

**Kakan Dey, Ph.D.**
**Assistant Professor**
Department of Civil & Environmental Engineering
West Virginia University
Office 647 ESB, Morgantown, WV 26506
Tel: (304)293-9952
Email: kakan.dey@mail.wvu.edu

**James Martin, Ph.D.**
Associate Professor
School of Computing, Clemson University,
211 McAdams Hall, Clemson, SC 29634, United States
Tel: 864-656-4529, Fax: +1-(864)-656-0145
Email: jmarty@clemson.edu

**ACKNOWLEDGEMENTS**

This material is based upon work supported by the USDOT Center for Connected Multimodal Mobility ($C^2M^2$) (Tier 1 University Transpiration Center) headquartered at Clemson University, Clemson, South Carolina, USA and the National Science Foundation (NSF) under U.S. Ignite Grant #1531127. Any opinions, findings, and conclusions or recommendations expressed in this material are those of the author(s) and do not necessarily reflect the views of the USDOT Center for Connected Multimodal Mobility ($C^2M^2$), NSF or USDOT, and the U.S. Government assumes no liability for the contents or use thereof.

**Will be published in the "2018 Transportation Research Board Conference Proceedings."**

**1.0 INTRODUCTION**
Cyber-Physical Systems (CPS) is an engineered system that seamlessly integrates computation, networking and physical devices. A Connected Vehicle (CV) system in which each vehicle can wirelessly communicate and share data with other vehicles or infrastructures (e.g., traffic signal, roadside unit), requires a cyber-physical system in order to predict traffic behavior through processing and aggregating data for improving safety and mobility, and reducing greenhouse gas emissions. A CPS with a centralized computing service cannot support real time CV applications due to the often unpredictable network latency (e.g., large latency), high data loss rate and expensive bandwidth, especially in a mobile network, such as in the CV environment *(1)*. Edge computing is a new concept for the CPS in which the resources for communication, computation, control, and storage are placed in a distributed manner at different edges *(2)*. This CPS system has the ability to reduce data loss and data delivery delay, and fulfill the high bandwidth requirements.

A queue downstream of a traveling vehicle is one of the major causes of rear-end collisions, and can disrupt traffic throughput by introducing shockwaves into the upstream traffic. About 28 percent crashes are rear-end collisions *(3)*. The queue prediction algorithm in a CV environment, presented in *(3)*, is based on the average separation distance between vehicles and the average speed of the surrounding vehicles in a traffic stream. This queue prediction algorithm uses Dedicated Short Range Communication (DSRC) to share vehicular information with other CVs every one tenth of a second. Typically, a queued state is determined by the average speed of the vehicles within the DSRC communication range as well as the average separation distance between each CV (i.e., subject vehicle) and the CV immediately in front of the subject vehicle. However, this algorithm detects queues depending on a specific threshold of average vehicle's speed and average separation distance. This queue detection algorithm is not applicable in mixed traffic scenarios, where non-connected vehicles can be between CVs and make the separation distance between CVs higher than the threshold. In the early days of CVs, data will be collected only from a limited number of CVs (i.e., low penetration rate), not from other vehicles (non-connected). Moreover, the data loss rate in the wireless CV environment contributes to the unavailability of data from the limited number of CVs *(4)*. Thus, it is very challenging to predict traffic behavior (e.g., queue prediction), which changes dynamically over time, with the limited CV data available in real-time.

The primary objective of this paper is to develop and evaluate the performance of an adaptive queue prediction application in an edge-centric CPS in mixed traffic. This algorithm is developed using a machine learning approach with a real-time feedback system. This adaptive queue prediction algorithm was evaluated using SUMO (i.e., Simulation of Urban Mobility) *(5)* and ns3 (Network Simulator 3) *(6)* to illustrate the efficacy of the adaptive queue prediction algorithm on a roadway network in Clemson, South Carolina, USA. The performance of the adaptive queue prediction application was measured in terms of queue detection accuracy with varying CV penetration levels and data loss rates. The analyses revealed that the adaptive queue prediction algorithm with feedback system outperforms without feedback system algorithm.

**2.0 METHOD**
In this section, we describe the details of an edge-centric CPS system for connected vehicle applications and the adaptive queue prediction algorithm in that edge-centric CPS.

**2.1 Edge-Centric Cyber-Physical System**
The edge-centric CPS consists of three edge levels: i) System Edge, ii) Fixed Edge, and iii) Mobile Edge. This hierarchical CPS architecture can address data processing complexity and scalability issues of any connected transportation systems. System Edge pertains to cloud-based services. A System Edge is a single end-point for a cluster of Fixed Edges. A Fixed Edge includes a general-purpose processor (i.e., edge device) and a DSRC based roadside unit (RSU). The Fixed Edge can communicate with the Mobile Edges (i.e., CVs) using DSRC and communicate with System Edge using Optical Fiber and/or WiFi. Fixed Edge can be extended to support video cameras and other sensing devices, such as weather sensors. CVs participating in the system will be acting as Mobile Edges, and be equipped with DSRC On Board Units (OBUs). Fixed Edges are connected through the backhauled network to a System Edge. DSRC OBUs can exchange data with the RSUs using DSRC communication.

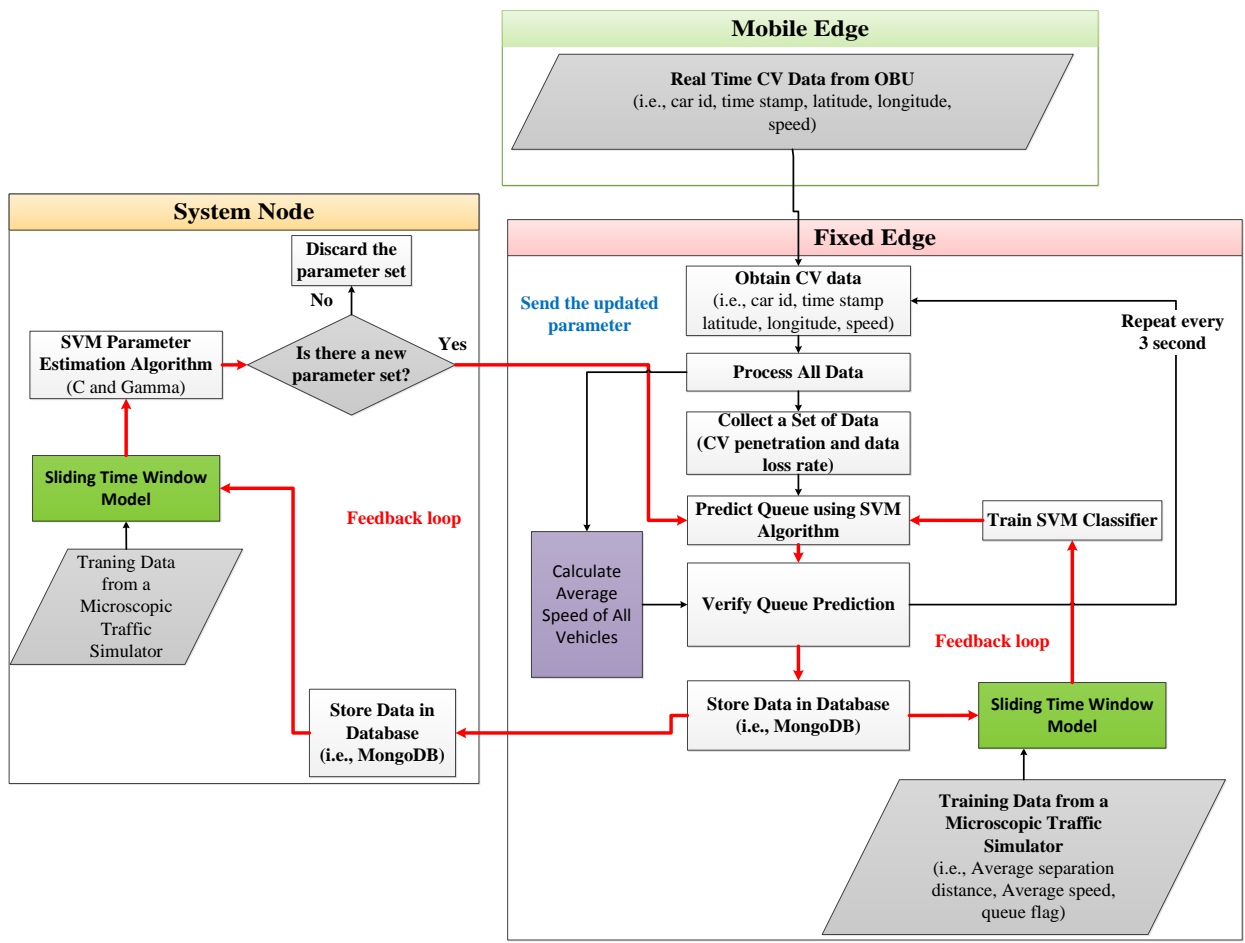

**FIGURE 1 Adaptive queue prediction in the edge-centric CPS.**

**2.2 Adaptive queue prediction algorithm using machine learning**
In this research, an adaptive queue prediction algorithm (as shown in Figure 1) is developed with a feedback system to determine if a CV is in a queued state or not. The queue prediction algorithm is deployed in different edge infrastructure components. Specifically, the machine learning (i.e., Support Vector Machine (SVM) is used in this study *(7)*) based queue prediction algorithm is deployed in the Fixed Edge layer and the parameter estimation process for the machine learning method is implemented in the System Edge to reduce the computational load at the Fixed Edges. Each Mobile Edge Node broadcasts Basic Safety Messages (BSMs) (e.g., the time stamp, latitude, longitude and speed of the vehicle) to the Fixed Edge nodes, within the DSRC coverage range, every one-tenth of a second. After receiving the raw BSMs from CVs, data will be processed and aggregated in the Fixed Edge, and then the aggregated data will be normalized for queue prediction. For training the machine-learning model, training data (i.e., speed, separation distance between vehicles) with label (queue or no queue) from a microscopic traffic simulator can be used initially. After traffic queue prediction for each time interval, traffic queue prediction is verified with ground truth data. For the verification of a predicted queue state, we used average speed of all the vehicles, which is used as ground truth data, on the corresponding road network used the simulation. The traffic state will be defined as a queue state if the average speed of all vehicles on the network is less than 5 mph *(3)* for the ground truth data. For traffic queue prediction for each time interval using SVM, we verify the traffic state (queue or no queue) and compare with the ground truth data. Traffic queue prediction data, verified with the simulation data, will be added to the training dataset for the next prediction time window, and support vector machine (SVM) parameters will be updated at the System Edge node (as shown in Figure 1). We store average speed, average separation distance and correct prediction results after verification (queue or no queue) in a database at Fixed Edge Node. The traffic state can be changed over time, and the evolving CV data stream needs to be added to the training data set for implementing a feedback loop in the system to improve queue prediction accuracy. This is how the queue prediction algorithm can adapt with the changes in different traffic conditions.

To update the training data set, we used a fixed sliding time window and a dynamic sliding time window method. According to the fixed sliding time window, the predicted labeled data will be stored for a certain time in the training dataset, and the same number of data rows will be removed from the oldest part of the training dataset. If we use the fixed sliding time window technique to update the training dataset, we may remove different patterns of labeled training data from the dataset. It will reduce the prediction accuracy, especially in evolving data streams. Because of this challenge, we introduce a dynamic sliding window technique *(8)* to update the training dataset. In our feedback system, we implement a parameter-free adaptive size sliding window method, which provides theoretical guarantees of the data pattern for the evolving real-time data streams *(8)*.

To evaluate the performance of the adaptive queue prediction algorithm, a baseline scenario of the queue prediction application without any feedback system was developed. We considered four different data loss rates with seven different CV penetration levels (as shown in Table 1) to evaluate the performance of the adaptive queue prediction algorithm.

**TABLE 1 Experimental Scenarios with and without Feedback System**

| Experimental Scenario | Data Loss Rate (%) | CV Penetration (%) |
|---|---|---|
| Scenario 1 | 2 | 10, 20, 30, 40, 50, 75, 100 |
| Scenario 2 | 4 | 10, 20, 30, 40, 50, 75, 100 |
| Scenario 3 | 8 | 10, 20, 30, 40, 50, 75, 100 |
| Scenario 4 | 16 | 10, 20, 30, 40, 50, 75, 100 |

## 3.0 FINDINGS

In Table 2, we compared the accuracy between the queue prediction algorithm with feedback system and without feedback system (using fixed sliding time window and dynamic sliding time window) with varying the CV penetration levels and data loss rates. It is evident that the accuracy of the queue prediction algorithm increases with increasing CV penetration levels for different data loss rates in both cases (i.e., with the feedback system and without the feedback system). The accuracy of the algorithm increases, as more data are available from a large number of CVs. Moreover, the rate of increase of queue prediction accuracy is higher for a CV penetration level of less than 50% than with a CV penetration level of 50% and above for each data loss rate scenario.

**TABLE 2 Comparison of Queue Prediction Accuracy Varying the CV Penetration Levels and Data Loss Rates between the without Feedback System and with feedback system (Fixed Sliding Time Window and Dynamic Sliding Time Window)**

| | *Queue prediction accuracy without feedback System* | | | |
|---|---|---|---|---|
| **CV Penetration (%)** | **Data loss (%)** | | | |
| | **2%** | **4%** | **8%** | **16%** |
| **10%** | 58% | 53% | 49% | 44% |
| **20%** | 71% | 68% | 64% | 60% |
| **30%** | 83% | 77% | 73% | 68% |
| **40%** | 86% | 80% | 78% | 73% |
| **50%** | 89% | 84% | 81% | 76% |
| **75%** | 92% | 88% | 83% | 80% |
| **100%** | 96% | 90% | 89% | 85% |
| | *Queue prediction accuracy with feedback System (Fixed Sliding Time Window)* | | | |
| **CV Penetration (%)** | **Data loss (%)** | | | |
| | **2%** | **4%** | **8%** | **16%** |
| **10%** | 64% | 60% | 54% | 50% |
| **20%** | 75% | 75% | 72% | 65% |
| **30%** | 88% | 80% | 77% | 75% |
| **40%** | 89% | 83% | 82% | 76% |
| **50%** | 93% | 87% | 84% | 82% |
| **75%** | 94% | 90% | 88% | 82% |
| **100%** | 97% | 92% | 90% | 88% |
| | *Queue prediction accuracy with feedback System (Dynamic Sliding Time Window)* | | | |
| **CV Penetration (%)** | **Data loss (%)** | | | |
| | **2%** | **4%** | **8%** | **16%** |
| **10%** | 67% | 62% | 57% | 53% |
| **20%** | 77% | 77% | 75% | 68% |
| **30%** | 92% | 83% | 80% | 78% |
| **40%** | 90% | 84% | 84% | 83% |
| **50%** | 94% | 90% | 87% | 82% |
| **75%** | 95% | 91% | 89% | 83% |
| **100%** | 98% | 93% | 92% | 90% |

The queue prediction algorithm with the feedback system using the fixed sliding time window and dynamic sliding time window were compared in terms of accuracy. Statistical analysis revealed that the overall queue prediction accuracy with the feedback system for both fixed sliding time and dynamic sliding time windows was significantly higher for each CV penetration level at a 95% confidence level compared to the queue prediction accuracy without the feedback system.

## 4.0 CONCLUSIONS

In the early days of CV environment, data will be collected from a limited number of CVs in mixed traffic. Moreover, the data loss rate in a wireless CV environment contributes to the unavailability of data from the limited number of CVs. In this study, a feedback system was developed using SVM so that the queue prediction accuracy can be improved in dynamic traffic conditions. We evaluated the accuracy of the queue prediction algorithm at different CV penetration levels and data loss rates in an integrated SUMO and ns-3 simulation. Our simulation analyses revealed that the queue prediction using the feedback system (i.e., adaptive queue prediction algorithm) has a higher accuracy compared to the queue prediction without a feedback system. The feedback system was developed using a fixed sliding time window and dynamic sliding time window. The unique feature of the dynamic time window is that verified queue prediction data sets with unique patterns are added from the evolving data streams in real-time depending on dynamic traffic conditions. Our analyses demonstrated that the accuracy for the dynamic sliding time window is higher than the fixed sliding time window for CV penetration level of less than 50%, and the performance of both fixed and dynamic sliding scenarios were very similar with a CV penetration level of 50% and above.